\documentclass[aps,twocolumn,showpacs,psfig,superscriptaddress,longbibliography]{revtex4-2} 

\usepackage{amsfonts}        
\usepackage{mathrsfs}        
\usepackage{amsmath}         
\usepackage{color}           
\usepackage{natbib}          
\usepackage{textcomp}        
\usepackage{graphicx}        
\usepackage{bm}              
\usepackage{amssymb}         

\usepackage{xspace}          
\usepackage{epstopdf}        
\usepackage{dcolumn}         
\usepackage{longtable}       
\usepackage{multirow}        
\usepackage[colorlinks=true, letterpaper=true, pdfstartview=FitV, linkcolor=blue, citecolor=blue, urlcolor=blue]{hyperref} 
\usepackage{braket}          
\usepackage{float}           
\usepackage{siunitx}         

\begin{document}

\setlength{\abovedisplayskip}{10pt plus 0pt minus 6pt}
\setlength{\belowdisplayskip}{10pt plus 0pt minus 6pt}

\title{\textit{Ab initio} time-dependent \textit{GW} approach for nonequilibrium exciton-phonon coupled dynamics across momentum space}

\author{Zhenfa Zheng}
\affiliation{Mork Family Department of Chemical Engineering and Materials Science, University of Southern 
California, Los Angeles, CA 90089, USA}

\author{Benran Zhang}
\affiliation{Mork Family Department of Chemical Engineering and Materials Science, University of Southern 
California, Los Angeles, CA 90089, USA}

\author{Jiawei Ruan}
\affiliation{Department of Physics, University of California at Berkeley, Berkeley, CA 94720, USA}
\affiliation{Materials Sciences Division, Lawrence Berkeley National Laboratory, Berkeley, California 94720, USA}

\author{Chih-En Hsu}
\affiliation{Mork Family Department of Chemical Engineering and Materials Science, University of Southern 
California, Los Angeles, CA 90089, USA}
\affiliation{Department of Physics, Tamkang University, Tamsui, New Taipei 251301, Taiwan}

\author{Zien Zhu}
\affiliation{Mork Family Department of Chemical Engineering and Materials Science, University of Southern 
California, Los Angeles, CA 90089, USA}

\author{Supavit Pokawanvit}
\affiliation{Department of Materials Science and Engineering, Stanford University, Stanford, CA 94305, USA}

\author{\\Felipe H. da Jornada}
\affiliation{Department of Materials Science and Engineering, Stanford University, Stanford, CA 94305, USA}

\author{Hung-Chung Hsueh}
\affiliation{Department of Physics, Tamkang University, Tamsui, New Taipei 251301, Taiwan}

\author{Ting Cao}
\affiliation{Department of Material Science and Engineering, University of Washington, Seattle, WA 98195, USA}

\author{Mauro Del Ben}
\affiliation{Applied Mathematics and Computational Research Division, Lawrence Berkeley National Laboratory, Berkeley, California 94720, USA}

\author{Yang-Hao Chan}
\email{yanghao@gate.sinica.edu.tw}
\affiliation{Institute of Atomic and Molecular Sciences, Academia Sinica, Taipei 10617, Taiwan}

\author{\\Steven G. Louie}
\email{sglouie@berkeley.edu}
\affiliation{Department of Physics, University of California at Berkeley, Berkeley, CA 94720, USA}
\affiliation{Materials Sciences Division, Lawrence Berkeley National Laboratory, Berkeley, California 94720, USA}

\author{Zhenglu Li}
\email{zhenglul@usc.edu}
\affiliation{Mork Family Department of Chemical Engineering and Materials Science, University of Southern 
California, Los Angeles, CA 90089, USA}

\begin{abstract}
The dynamics of optical excitations in materials generally involves intertwined electron-hole ($e$-$h$) and electron-phonon ($e$-ph) interactions out of equilibrium.
However, a full theoretical description of such nonequilibrium dynamics requires a systematic treatment of the coherent excitonic excitations and exciton-phonon interactions across the entire crystal momentum space in real time, which remains a major challenge and out of reach for first-principles approaches.
Here, we present a new \textit{ab initio} time-dependent adiabatic $GW$ methodology that incorporates full finite-momentum $e$-$h$ and $e$-ph couplings, enabling real-time simulations of the coherently coupled exciton-phonon dynamics.
The excitonic excitations are naturally described by the equation of motion of the interacting single-particle density matrix, whereas their couplings to phonons are formulated within a linear-response framework, hence the simulations can be efficiently carried out within a primitive unit cell.
We demonstrate the capabilities of this new approach by investigating the direct-to-indirect exciton transitions in monolayer WSe$_2$ in a pump-probe setup of time-resolved and angle-resolved photoemission spectroscopy. 
Our results reveal that the phonon-mediated ultrafast intervalley dynamics of excitons of this system is within $\sim$0.5~ps, manifested as in-gap photoemission intensity transfer from the \textit{K}-valley to the \textit{Q}-valley.
This work establishes a comprehensive and practical nonequilibrium Green's function framework for accurately simulating nonequilibrium and coherent excitations involving coupled excitons and phonons from first principles.
\end{abstract} 

\maketitle

\textit{Introduction.}
Light-matter interaction is a major driver of nonequilibrium dynamics of materials.
On ultrafast time scales, e.g., $\sim$fs to $\sim$ps, optical excitations, such as excitons, couple to the lattice vibrations coherently, shaping a wide range of phenomena including photo-induced structural phase transitions and vibronic energy transfer~\cite{deLaTorre2021,Thouin2019,scholes2017using,engel2007evidence,Madeo2020}.
A microscopic understanding of these phenomena in real materials therefore necessitates a comprehensive first-principles approach, posing a major challenge where the nonequilibrium dynamical nature and the coherent coupling between excitons and phonons must be accounted for in a unified many-body framework~\cite{Attaccalite2011,Sangalli2018,Chen2020,Antonius2022,Chan2021,Jiang2021,Perfetto2022,Chen2022,Stefanucci2023,zheng2023ab,chan2024exciton,Dai2024,bai2024ab,stefanucci2026unified}.
Particularly, a full description of the coherent exciton-phonon coupled dynamics requires the incorporation of the electron-hole ($e$-$h$) and electron-phonon ($e$-ph) interactions across the full Brillouin zone (BZ), which presently remains inaccessible from first principles.

Many-body perturbation theory provides the systematically formal~\cite{Hedin1965,hedin1970effects,strinati1988application} and first-principles~\cite{hybertsen1985first,Hybertsen1986,Onida2002,louie2021discovering} frameworks for excited-state phenomena.
Through decades of development and application, the $GW$ plus Bethe-Salpeter equation ($GW$-BSE) approach has demonstrated excellent accuracy in describing equilibrium excitonic and optical properties across a wide range of materials~\cite{hedin1970effects,strinati1988application,hybertsen1985first,rohlfing1998electron,albrecht1998ab,benedict1998optical,Rohlfing2000}.
Within the framework of nonequilibrium Green's functions, a time-dependent adiabatic $GW$ (TD-a$GW$) approach has recently been developed~\cite{Chan2021,Attaccalite2011,rocca2010ab,Chan2023}, which to some extent can be viewed as a time-dependent generalization of $GW$-BSE.
The TD-a$GW$ method is equivalent to an interacting single-particle density matrix formalism~\cite{Chan2021,Attaccalite2011,rocca2010ab} and provides first-principles simulations of exciton dynamics both in and out of equilibrium, with applications to pump-probe responses~\cite{hu2023excitonic}, nonlinear optics~\cite{Chan2021,hu2023light,chang2024many}, and exciton-Floquet engineering~\cite{Chan2023,Pareek2026}. 
However, it assumes clamped nuclei and omits phonons entirely, which are precisely the dominant degrees of freedom driving intervalley exciton dynamics in transition-metal dichalcogenides and finite-momentum exciton transfer in general~\cite{Madeo2020,Chen2022,chan2024exciton}.
Incorporating exciton-phonon coupling while preserving coherence is highly nontrivial, as the density matrix in the Bloch band state basis becomes non-diagonal in crystal momentum, acquiring finite-momentum coherence across the BZ and intertwining all momentum components concurrently.

\begin{figure}[!t]
\centering
\includegraphics[width=1.0\columnwidth]{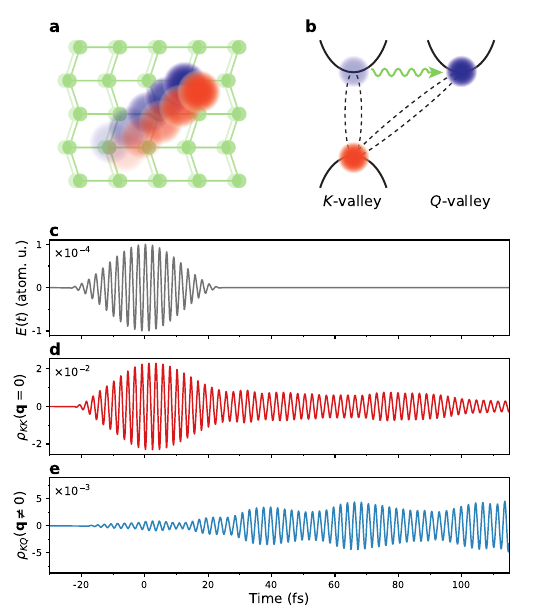}
\caption{Characteristics of nonequilibrium exciton-phonon coupled dynamics. \textbf{a}. Schematic illustration of exciton dynamics mediated by phonons, where the orange (blue) disk represents quasi-hole (quasi-electron) in an exciton, and the green background represents a vibrating lattice. \textbf{b}. Schematic illustration of phonon-mediated intervalley coupling showing a direct exciton at the \textit{K}-valley scattering to an indirect exciton at the \textit{Q}-valley in monolayer WSe$_2$. The green wavy arrow denotes a phonon and the dashed lines indicate bound electron-hole pairs. \textbf{c}. Electric field $E(t)$ of a pump pulse of duration of 50~fs. \textbf{d}. Zero-momentum density-matrix coherence factor $\rho_{KK}(\textbf{q}=0)$ from summing over two equivalent \textit{K}-valleys as defined in the text, created  by the pump field. \textbf{e}. Finite-momentum coherence factor $\rho_{KQ}(\textbf{q}\neq 0)$ from summing over all equivalent $\mathbf{q} = \textit{Q} -\textit{K}$, which is built up gradually during the pulse and after the pulse is off.}
\label{fig1}
\end{figure}

In this work, \textit{for the first time}, we develop a new \textit{ab initio} method that incorporates the coherent $e$-$h$ and $e$-ph couplings across the full BZ building on top of the TD-a$GW$ framework, denoted as TD-a$GW$-ph.
The TD-a$GW$-ph method overcomes the aforementioned challenges and enables a comprehensive many-body description of the nonequilibrium exciton dynamics and its coherent coupling to phonons (Fig.~\ref{fig1}\textbf{a}) from first principles.
Through a linear-response expansion of the density operator and $GW$ quasiparticle Hamiltonian with respect to the phonon perturbations, the coherent couplings across the crystal momentum space, i.e., the full BZ, can be compactly formulated and efficiently computed within a primitive unit cell through coupled equations of motion (EOMs).
We apply TD-a$GW$-ph to study the exciton dynamics in monolayer WSe$_2$ observed in time-resolved and angle-resolved photoemission spectroscopy (tr-ARPES)~\cite{Madeo2020}, where direct bright excitons (from intravalley \textit{K}-to-\textit{K} \textit{e}-\textit{h} excitations) are excited in a pump-probe setup and then rapidly transferred to indirect dark excitons (from intervalley \textit{K}-to-\textit{Q} \textit{e}-\textit{h} excitations) (Fig.~\ref{fig1}\textbf{b}).
Our time-dependent simulations directly and unambiguously reveal the phonon-mediated direct-to-indirect exciton transition mechanism and the transfer of the ARPES intensity (i.e., the single-particle spectral weight) across the energy and momentum space on an ultrafast time scale within $\sim$0.5 ps.

\textit{Method.}
Existing \textit{ab initio} TD-a$GW$ methods are developed assuming clamped nuclei~\cite{Attaccalite2011,Chan2021}, whereas in reality, phonon effects exist ubiquitously even at zero temperature via zero-point motion~\cite{Giustino2017}. 
To generalize to and derive the TD-a$GW$-ph formalism, we explicitly track the \textit{time-dependent} atomic displacement $\textbf{u}_{\kappa l}(t)$ of the $\kappa$-th atom in the $l$-th unit cell for the whole crystal.
In the TD-a$GW$ framework, the EOM propagates the interacting single-particle density operator $\hat{\rho}$ (equivalent to single-particle equal-time lesser Green's function) in real time following the Liouville-von Neumann equation,
\begin{equation} \label{td-agw EOM}
    i\hbar \frac{\partial}{\partial t} \hat{\rho} (t; \{\textbf{u}_{\kappa l}\} ) = \left[ \hat{H}^{\text{a}GW}(t;\{\textbf{u}_{\kappa l}\} ) , \ \hat{\rho}(t;\{\textbf{u}_{\kappa l}\} )  \right] .
\end{equation}
The TD-a$GW$ Hamiltonian is taken to be,
\begin{equation}
\begin{split}
     & \hat{H}^{\text{a}GW}(t;\{\textbf{u}_{\kappa l}\} )  \\
   = & \hat{h}(t;\{\textbf{u}_{\kappa l}\} )  + \hat{U}^{\text{ext}}(t) + \delta \hat{V}^{e\text{-}h}(t;\{\textbf{u}_{\kappa l}\} ) ,
\end{split}
\end{equation}
where $\hat{h}$ is the equilibrium single-particle Hamiltonian at the $GW$ level. Upon application of the external light field $\hat{U}^\text{ext} (t) = -e\textbf{E}(t)\cdot \hat{\textbf{r}}$, we separate and explicitly denote the variation in the $GW$ Hamiltonian induced by the field as $\delta \hat{V}^{e\text{-}h}$. Hence $\hat{h}$ only accounts for the field-unperturbed part of the $GW$ quasiparticle Hamiltonian. We also emphasize here that $\delta \hat{V}^{e\text{-}h}$ contains the field-induced time variations of both the Hartree and GW self-energy terms in the GW formalism.
We label it as the $\delta \hat{V}^{e\text{-}h}$ term since it corresponds to the $GW$-BSE $e$-$h$ interaction~\cite{Rohlfing2000,Onida2002,Chan2021,Attaccalite2011} and captures excitonic effects in response to an external light field $\hat{U}^\text{ext}$ with a time-dependent generalization. 

To initialize the density operator for the time evolution, we introduce an \textit{ansatz} in the Born-Oppenheimer adiabatic limit. 
Without the external field (i.e., $\hat{U}^\text{ext} = 0$), at a given $t$, the $GW$ Hamiltonian $\hat{h}(t;\{\textbf{u}_{\kappa l}\} )$ can be diagonalized to obtain the corresponding quasiparticle eigenstates $ \tilde{\psi}_{n\textbf{k}}(t;\{\textbf{u}_{\kappa l}\})$  ($n$: band index; \textbf{k}: electron wavevector), which form a set of states to conveniently construct the field-unperturbed density operator,
\begin{equation} \label{rho tilde}
    \hat{\tilde{\rho}} (t) = \sum_{n\textbf{k}} f_{n\textbf{k}}(t) \ket{\tilde{\psi}_{n\textbf{k}}(t)} \bra{\tilde{\psi}_{n\textbf{k}}(t)} ,
\end{equation}
where $f_{n\textbf{k}}$ is the Fermi occupation.

In developing TD-a$GW$-ph, we introduce a linear-response formalism in the phonon perturbations (atomic displacements), allowing for evolving the EOMs within the primitive unit cell. 
We denote the first-order phonon-perturbation operator (the displacements of the atoms in terms of phonon operators $\hat{a}_{\textbf{q}\nu}$ and $\hat{a}^\dagger_{\textbf{q}\nu}$)~\cite{Giustino2017,Baroni2001},
\begin{equation}
    \Delta = \frac{1}{\sqrt{ N_\textbf{q}}} \sum_{\textbf{q}\nu} (\hat{a}_{\textbf{q}\nu} + \hat{a}^\dagger_{-\textbf{q}\nu}) \Delta_{\textbf{q}\nu},
\end{equation}
where the individual mode-resolved (\textbf{q}: phonon wavevector, $\nu$: branch index) perturbation operator component carries a crystal momentum $\textbf{q}$ and has the form,
\begin{equation}
    \Delta_{\textbf{q}\nu} = \sqrt{\frac{\hbar}{2\omega_{\textbf{q}\nu}}} \sum_{\kappa\alpha} \frac{1}{\sqrt{M_\kappa}} e_{\kappa\alpha,\nu}(\textbf{q}) \sum_l^{N_\textbf{q}} e^{i\textbf{q}\cdot\textbf{R}_l} \frac{\partial}{\partial u_{\kappa\alpha l}},
\end{equation}
with phonon frequency $\omega_{\textbf{q}\nu}$ and corresponding eigenvector $\textbf{e}_\nu(\textbf{q})$, $u_{\kappa\alpha l}$ labeling the $\alpha$ Cartesian component of $\textbf{u}_{\kappa l}$, $M_\kappa$ the atomic mass, and $\textbf{R}_l$ a lattice vector.
The phonon-perturbed $GW$ Hamiltonian can be built in the following form,
\begin{equation} \label{perturbed h}
    \hat{h}(t;\{\textbf{u}_{\kappa l}\} ) = \hat{{h}}(\{\textbf{u}_{\kappa l} = 0\} ) + \hat{h}^{e\text{-ph}},
\end{equation}
with the $e$-ph coupling Hamiltonian $\hat{h}^{e\text{-ph}} = \Delta \hat{h}$ written in expansion as,
\begin{equation} \label{heph}
\begin{split}
     \hat{h}^{e\text{-ph}} = & \frac{1}{\sqrt{ N_\textbf{q}}} \sum_{mn\textbf{k}} \sum_{\textbf{q}\nu} (\hat{a}_{\textbf{q}\nu} + \hat{a}^\dagger_{-\textbf{q}\nu}) \\
     & \times
         g_{mn\nu}(\textbf{k}, \textbf{q}) \ket{\bar\psi_{m\textbf{k}+\textbf{q}}}\bra{\bar\psi_{n\textbf{k}}},
\end{split}
\end{equation}
where $\bar{\psi}_{n\textbf{k}}(\{\textbf{u}_{\kappa l} = 0\})$ denotes the Bloch state wavefunctions at the equilibrium positions, and $g_{mn\nu}(\textbf{k}, \textbf{q})$ represents the $e$-ph matrix elements (see Supplementary Information (SI)).

\begin{figure*}[!t]
\centering
\includegraphics[width=1.0\textwidth]{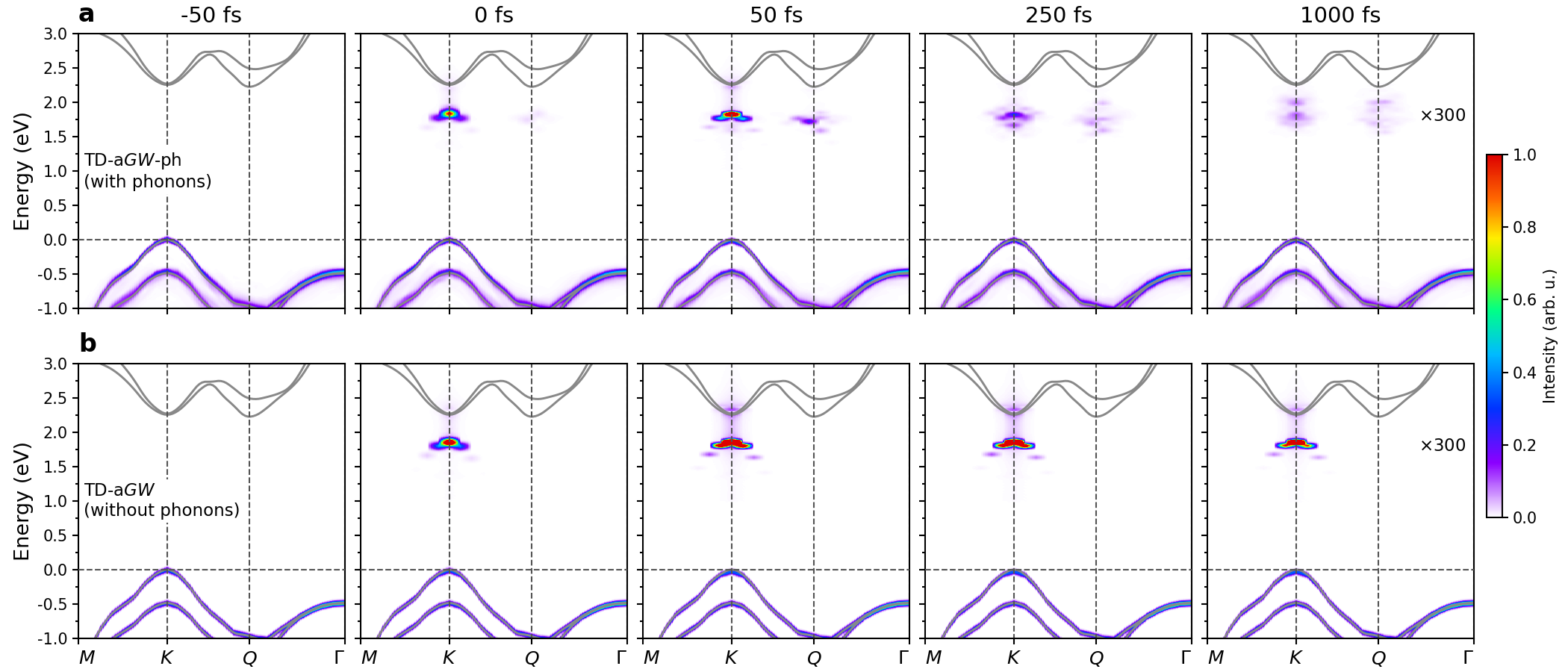}
\caption{Theoretical simulated tr-ARPES spectra at 100~K with pump pulse given in Fig.~\ref{fig1}\textbf{c}.  \textbf{a}. Time evolution of the spectral function with phonon coupling included and evaluated at $t = -50, 0, 50, 250, 1000$~fs using the TD-a$GW$-ph method. \textbf{b}. Simulations as in \textbf{a} but without coupling to phonons using the TD-a$GW$ method. Gray lines denote the equilibrium $GW$ band structure. Spectral intensities above 1.0~eV are multiplied by a factor of 300. A dephasing time $\tau_{\text{deph}}=2.2~\text{ps}$ is included (see Fig.~\ref{fig3}\textbf{b}).}
\label{fig2}
\end{figure*}

To proceed forward, we use the Bloch wavefunctions at equilibrium positions $\{\ket{\bar{\psi}_{n\textbf{k}}}\}$ as the basis set to evaluate Eq.~\eqref{td-agw EOM}, with the convenience of this choice to be discussed below.
The inclusion of phonons allows for finite-momentum  ($\textbf{q}\neq 0$) density-matrix elements
$\rho_{m\textbf{k}+\textbf{q}, n\textbf{k}} =\braket{\bar{\psi}_{m\textbf{k}+\textbf{q}} |\hat{\rho}|\bar{\psi}_{n\textbf{k}}}$
between Bloch band states at different wavevectors to be non-zero.
By further expanding the commutator in Eq.~\eqref{td-agw EOM} and inserting the identity $\hat{I} = \sum_{l\textbf{k}'} \ket{\bar{\psi}_{l\textbf{k}'}}\bra{\bar{\psi}_{l\textbf{k}'}}$ in between operators in a product, we arrive at the central EOM of the TD-a$GW$-ph method,
\begin{widetext}
\begin{equation} \label{central eom}
\begin{split}
    i\hbar \frac{\partial}{\partial t} \rho_{m\textbf{k}+\textbf{q}, n\textbf{k}}(t)
    & = \left( \varepsilon_{m\textbf{k}+\textbf{q}} - \varepsilon_{n\textbf{k}} \right) \rho_{m\textbf{k}+\textbf{q}, n\textbf{k}}(t)
    + \sum_l \biggr( U^\text{ext}_{m\textbf{k}+\textbf{q}, l\textbf{k}+\textbf{q}}(t) \rho_{l\textbf{k}+\textbf{q}, n\textbf{k}}(t) 
    - \rho_{m\textbf{k}+\textbf{q}, l\textbf{k}}(t) U^\text{ext}_{l\textbf{k}, n\textbf{k}}(t) \biggr) \\
    & \quad + \sum_{l\textbf{q}'} \biggr( M_{m\textbf{k}+\textbf{q}, l\textbf{k}+\textbf{q}'}(t) \rho_{l\textbf{k}+\textbf{q}', n\textbf{k}}(t)
    - \rho_{m\textbf{k}+\textbf{q}, l\textbf{k}+\textbf{q}'}(t) M_{l\textbf{k}+\textbf{q}', n\textbf{k}}(t)
    \biggr),
\end{split}
\end{equation}
\end{widetext}
where $\varepsilon_{n\textbf{k}}$ is the $GW$ band energy, and the interaction matrix element is defined as,
\begin{equation} \label{mint}
\begin{split}
    & M_{m\textbf{k}+\textbf{q},n\textbf{k}}(t) \\
    = & \delta V^{e\text{-}h}_{m\textbf{k}+\textbf{q}, n\textbf{k}}(t) 
    + \frac{1}{\sqrt{N_{\textbf{q}}}}\sum_\nu g_{mn\nu}(\textbf{k}, \textbf{q})
    \braket{\hat{a}_{\textbf{q}\nu} + \hat{a}^\dagger_{-\textbf{q}\nu}}.
\end{split}
\end{equation}
Here, $M_{m\textbf{k}+\textbf{q},n\textbf{k}}$ consists of $e$-$h$ and $e$-ph interactions \textit{coherently} in both the gauge of the Bloch states (a technical requirement, see SI) and the time dependence in the dynamics of the $e$-$h$ pairs and phonon vibrations (a physical consequence).
The latter becomes clear with the classical replacement of the phonon ladder operators~\cite{zacharias2016one},
\begin{equation}
    \braket{\hat{a}_{\textbf{q}\nu} + \hat{a}^\dagger_{-\textbf{q}\nu}} \rightarrow \sqrt{4 n_{\textbf{q}\nu} + 2} \cos(\omega_{\textbf{q}\nu}t + \phi_{\textbf{q}\nu}),
\end{equation}
where $n_{\textbf{q}\nu}$ is the Boson occupation and $\phi_{\textbf{q}\nu}$ is a random mode-dependent (initial) phase.
In this work, since we mostly focus on the ultrafast dynamics ($t \lesssim 1$ ps),  we assume background phonons at a fixed temperature, and neglect the electronic feedback on phonon EOMs.
The generalization to include phonon EOMs into TD-a$GW$-ph is relatively straightforward (as achieved in the case of zone-center $\textbf{q}=0$ phonon~\cite{stefanucci2024semiconductor,Chan2025raman}) and a subject for future work, which holds promise to simulate light-structure coupling phenomena such as the formation of self-trapped excitons~\cite{Dai2024,bai2024ab}.
Additionally, our classical treatment of phonons does not distinguish the asymmetry between the quantum phonon absorption and emission processes~\cite{zhang2024phonon}, which may overestimate phonon-absorption channels and underestimate phonon-emission channels at low temperature.
The $e$-$h$ interaction term $\delta \hat{V}^{e\text{-}h}$ contains the light-induced variations in the Hartree potential and the $GW$ self-energy (in practice, within the static-screening approximation)~\cite{Chan2021}, $\delta \hat{V}^{e\text{-}h} = \delta \hat{V}^\text{H} + \delta \hat{\Sigma} $, which lead to the exchange and direct $e$-$h$ interaction kernels in the $GW$-BSE formalism, respectively. 
Specifically, the $\delta \hat{V}^{e\text{-}h}$ term can be written in terms of the matrix elements of the total $e$-$h$ kernel $\hat{K}^{e\text{-}h}$ (including exchange and direct kernels, see SI)~\cite{qiu2015nonanalyticity}, as well as a functional of the density matrix,
\begin{equation} \label{veh}
\begin{split}
    & \delta V^{e\text{-}h}_{m\textbf{k}+\textbf{q}, n\textbf{k}}(t) \\
    = & \sum_{jl\textbf{k}'} K^{e\text{-}h}_{m\textbf{k}+\textbf{q},n\textbf{k}; j\textbf{k}'+\textbf{q}, l\textbf{k}'}
    \Big( \rho_{j\textbf{k}'+\textbf{q}, l\textbf{k}'}(t)
    - \tilde{\rho}_{j\textbf{k}'+\textbf{q}, l\textbf{k}'}(t)
    \Big).
\end{split}
\end{equation}
Note that the light-induced variation needs to be explicitly treated by removing the external field-unperturbed part from the total density matrix. 
Here, we take the adiabatic ansatz $\hat{\tilde{\rho}}$ in Eq.~\eqref{rho tilde} as the field-unperturbed density matrix.
In arriving at Eq.~\eqref{veh}, we neglect the phonon perturbation effects in the $e$-$h$ interaction kernel, which is a commonly adopted and appropriately justified approximation, $\partial \hat{K}^{e-h} / \partial u_{\kappa\alpha l} = 0$, in the exciton-phonon coupling studies~\cite{Chen2020,Antonius2022}. 
The choice of the basis of $\{\ket{\bar{\psi}_{n\textbf{k}}}\}$ allows us, in the evaluation of Eq.~\eqref{mint} and Eq.~\eqref{veh}, to directly use the $e$-ph coupling and $e$-$h$ kernel matrix elements computed at the equilibrium atomic positions, which are readily available from density-functional perturbation theory~\cite{Baroni2001} or $GW$ perturbation theory~\cite{Li2019GWPT,li2024electron}, and the $GW$-BSE calculations, respectively.
Because of the nearly zero-momentum nature of photons, the matrix elements of the light field $\hat{U}^\text{ext}$ may be approximated as diagonal in the wavevectors in Eq.~\eqref{central eom}.

Our development of the TD-a$GW$-ph method aims at describing exciton-phonon dynamics in semiconductors and insulators. Therefore, in this work, we further reduce the complexity of the computation through applying the Tamm-Dancoff approximation (TDA)~\cite{Rohlfing2000,Onida2002}, which only retains $e$-$h$ kernel matrix elements of the excitation type --- i.e., by keeping only the $cv \leftrightarrow cv$ and $vc \leftrightarrow vc$ transitions ($c$: conduction band index; $v$: valence band index) and neglecting all other transitions including those between $vc$ and $cv$ and those involving $cc$ and $vv$. 
Consequently, within this approximation, the $e$-ph coupling then should not mix the conduction and valence manifolds, i.e., by keeping only $c \leftrightarrow c$ and $v \leftrightarrow v$ scatterings in $g_{mn\nu}(\textbf{k}, \textbf{q})$, to preserve the TDA subspace for excitons. 
In this way, the interaction landscape in the TD-a$GW$-ph approach resembles the existing exciton-phonon coupling formalism~\cite{Chen2020,Antonius2022}, with the time-dependent generalization to describe nonequilibrium exciton-phonon coupled dynamics. 
Under TDA, along with expanding the phonon-perturbed wavefunctions using the first-order perturbation theory as $\tilde{\psi}_{n\textbf{k}}(t;\{\textbf{u}_{\kappa l}\}) = \bar{\psi}_{n\textbf{k}}(\{\textbf{u}_{\kappa l} = 0\}) + \Delta \psi_{n\textbf{k}}(t;\{\textbf{u}_{\kappa l}\})$, the adiabatic ansatz of the field-unperturbed density matrix takes a simple time-independent form,
\begin{equation}
    \tilde{\rho}^\text{TDA}_{m\textbf{k}+\textbf{q}, n\textbf{k}}(t) = f_{n\textbf{k}}\delta_{mn}\delta_{\textbf{q}\Gamma} ,
\end{equation}
where $f_{n\textbf{k}}$ stays constant for materials with sizable band gaps that remain open upon phonon perturbations.

The TD-a$GW$-ph dynamics is much more challenging to compute, compared with the TD-a$GW$ approach.
In TD-a$GW$, the density matrix is diagonal in wavevectors (i.e., $\textbf{q} = 0$ only), whereas in TD-a$GW$-ph, there are $N_\textbf{q}$ times more EOMs evolving together, capturing the finite-momentum coherence. 
Moreover, as shown in Eq.~\eqref{central eom}, each $\textbf{q}$-component of the density matrix communicates with all $N_\textbf{q}$ components at every time step, collectively contributing an additional $O(N_\textbf{q}^2)$ complexity factor in TD-a$GW$-ph \textit{vs} TD-a$GW$.
We have implemented the above TD-a$GW$-ph formalism in the \textsc{BerkeleyGW} software package~\cite{Deslippe2012,delben2020accelerating,zhang2025advancing} (see SI).

\begin{figure*}[!t]
\centering
\includegraphics[width=1.0\textwidth]{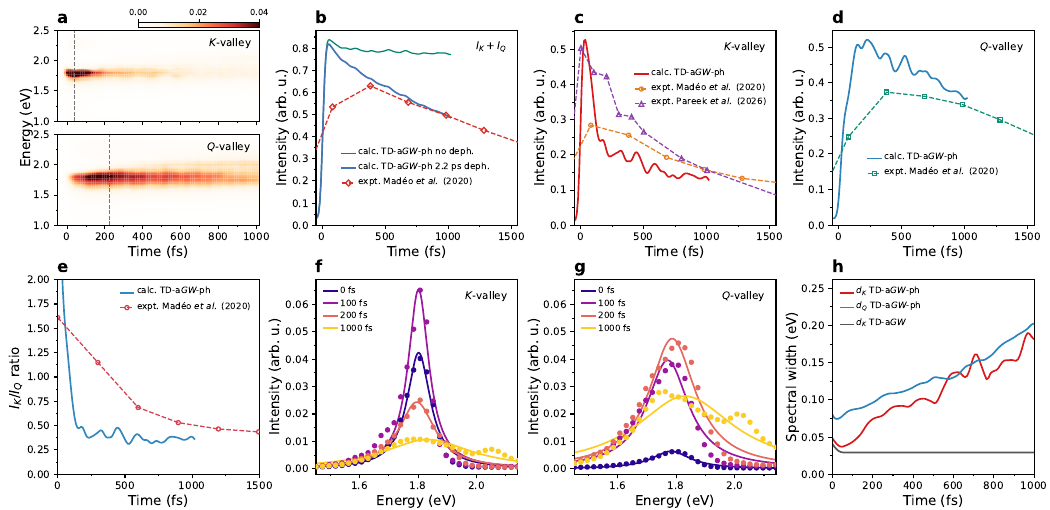}
\caption{Intervalley exciton transfer dynamics in monolayer WSe$_2$. 
\textbf{a}. Time-energy resolved spectral intensity $I(E,t)$ of the \textit{K}- and \textit{Q}-valley, which is related to tr-ARPES intensity, with the \textbf{k}-distribution of the intensity integrated in the respective valley within a circular region of radius of 0.23~\AA$^{-1}$ (summing over equivalent valleys). The intensity peaks at 38 fs at \textit{K}-valley and 222 fs at \textit{Q}-valley, indicated by the vertical dashed lines. 
\textbf{b}. Sum of the intensities integrated over the energy window 1.0--2.5~eV for the \textit{K}-valley $I_K$ and \textit{Q}-valley $I_Q$ from calculation and experiment~\cite{Madeo2020}. The green line denotes the full intensity without dephasing, and the blue line denotes the results computed with a dephasing time $\tau_\text{deph} = 2.2$~ps, fitted from the experimental decay.
This dephasing time is included in Fig.~\ref{fig2} and all subsequent panels in Fig.~\ref{fig3}. 
The experimental data~\cite{Madeo2020} is normalized to the theoretical $I_K + I_Q$ (with dephasing) at 1000 fs.  
\textbf{c}. Energy-integrated intensity $I_K$ at the \textit{K}-valleys from calculation and experiments~\cite{Madeo2020,Pareek2026}.  The purple experimental data set~\cite{Pareek2026} is for a similar material WS$_2$ measured more recently, and is normalized to the orange data set of WSe$_2$~\cite{Madeo2020} at 1000 fs.
\textbf{d}. Energy-integrated intensity $I_Q$ at the \textit{Q}-valleys from calculation and experiment~\cite{Madeo2020}.
In \textbf{c} and \textbf{d}, the intensities include contributions from all degenerate valleys.
\textbf{e}. Intensity ratio $I_K/I_Q$ from calculation and experiment~\cite{Madeo2020}. The ratio decreases from $>$1.5 to a quasi-equilibrium of $\sim$0.5 beyond $\sim$500~fs.
\textbf{f}. Energy-dependent intensity line cuts from \textbf{a} at different time snapshots in the \textit{K}-valley, fitted to Lorentzian functions. 
\textbf{g}. Similar to \textbf{f} but for \textit{Q}-valley.
\textbf{h}. Half-width-at-half-maximum spectral width $d(t)$ of the Lorentzian fitting from \textbf{f} and \textbf{g} for the \textit{K}-valley ($d_K$) and \textit{Q}-valley ($d_Q$) using the TD-a$GW$-ph method, showing a clear increasing trend as a function of time induced by classical phonons. For comparison, $d_K$ without phonon effects using the TD-a$GW$ method is also presented, displaying a constant spectral width after the pump pulse ended.
}
\label{fig3}
\end{figure*}

\textit{Results.}
Monolayer WSe$_2$ is a prototypical two-dimensional (2D) transition-metal dichalcogenide in which spin-orbit coupling splits the top of the valence band, placing the lowest-energy bright exciton at the \textit{K}-valleys (double degeneracy in the BZ)~\cite{Xiao2012,cao2012valley,Madeo2020}. The \textit{Q}-valleys (sextuple degeneracy in the BZ), located between $\Gamma$ and \textit{K}, host conduction band minima nearly degenerate with the \textit{K}-valley conduction band minima (see SI), making WSe$_2$ an ideal platform for phonon-mediated direct-to-indirect (bright-to-dark) exciton transfer as has been directly observed in tr-ARPES~\cite{Madeo2020,Pareek2026,wu2026excitons}.
We use the new TD-a$GW$-ph method developed in this work to simulate the exciton dynamics of monolayer WSe$_2$. 
We apply a pump pulse tuned in resonance with the lowest intravalley bright exciton (1.85~eV) with a cosine-squared 50~fs envelope (Fig.~\ref{fig1}\textbf{c}) (see SI).
A dephasing time constant of $\tau_{\text{deth}}=2.2~\text{ps}$ is included in the dynamics simulations (see further discussions below and in SI).
We track two representative types of density-matrix coherence factors (summing over selected components of the density matrix) between the lowest conduction band ($c_1$) and the highest valence band ($v_1$): the zero-momentum coherence factor situated at \textit{K}-valleys, $\rho_{\textit{K}\textit{K}} = \sum_{\textbf{k}= K, -K} \left( \rho_{c_1\textbf{k},v_1\textbf{k}} + \rho_{v_1\textbf{k},c_1\textbf{k}} \right)$, and the finite-momentum coherence factor $\rho_{\textit{K}\textit{Q}} = \sum_{\textbf{k}= K, -K}\sum_{\textbf{q} \in \{Q-\textbf{k}\}_\text{n.n.}} \left( \rho_{c_1\textbf{k}+\textbf{q},v_1\textbf{k}}  + \rho_{v_1\textbf{k},c_1\textbf{k}+\textbf{q}} \right)$ connecting \textit{K}-valley holes to the \textit{Q}-valley electrons via \textbf{q} as the wavevector difference between nearest-neighbor \textit{Q} and \textit{K} points. In the beginning of the dynamics, only $\rho_\textit{KK}$ with zero momentum ($\textbf{q}=0$) couples directly to the light field, responding instantaneously to the pulse $E(t)$ (Fig.~\ref{fig1}\textbf{d}). 
The finite-momentum ($\textbf{q}\neq0$) coherence factor $\rho_\textit{KQ}$, by contrast, cannot be generated optically, but builds up steadily (Fig.~\ref{fig1}\textbf{e}) via phonon-mediated coupling between the \textit{K}- and \textit{Q}-valleys. These finite-momentum density-matrix components constitute the central new physics from TD-a$GW$-ph, which are absent in the clamped-nuclei TD-a$GW$ approach.

Fig.~\ref{fig2}\textbf{a} shows the calculated tr-ARPES intensity map $I_{\mathbf{k}}(E,t)$~\cite{Chan2023,freericks2009theoretical,stefanucci2013nonequilibrium} along the \textit{M}-\textit{K}-\textit{Q}-$\Gamma$ path using the TD-a$GW$-ph approach (see SI), demonstrating the intervalley exciton transfer dynamics in monolayer WSe$_2$. At $t = -50$~fs before the pump is applied, only the valence bands are occupied and no in-gap exciton-derived signal is present. At $t = 0$~fs, the resonant pump generates a bright exciton at the \textit{K}-valley, appearing as a downward bending photoemission feature below the single-particle conduction band minimum, while the \textit{Q}-valley starts to show non-negligible intensity. As the system evolves further, the \textit{K}-valley intensity becomes stronger at 50~fs and then drops at 250~fs. In the meantime, a pronounced spectral feature builds up at the \textit{Q}-valley, directly demonstrating the intervalley exciton transfer dynamics across the crystal momentum space. By $t = 1000$~fs, the distribution of the excitonic weight across both valleys becomes steady and broadened by the phonon scattering.
As a direct comparison, we perform standard TD-a$GW$ calculations as shown in Fig.~\ref{fig2}\textbf{b}, excluding $e$-ph coupling and phonon effects. 
As a result, the \textit{Q}-valley stays completely dark, and the intensity at the \textit{K}-valley stays within the same valley, unambiguously establishing the indispensable role of phonons for the intervalley exciton transfer.

\textit{Discussions.}
To quantify the intervalley transfer, we plot the time-energy resolved spectral intensity $I(E, t)$, with the \textbf{k}-dependence integrated within a circular region of radius of 0.23~\AA$^{-1}$ in respective \textit{K}- and \textit{Q}-valleys (summing over equivalent valleys), as shown in Fig.~\ref{fig3}\textbf{a}.
The \textit{K}-valley signal emerges instantaneously with the pump with a peak at 38 fs, and decays within a few hundred femtoseconds, while the \textit{Q}-valley signal builds up with a clear time delay with a peak at 222 fs.  The $\sim$180 fs separation between the two peaks directly characterizes the phonon-mediated intervalley transfer timescale. 
In the next step of the analysis, we further integrate the intensities over the energy window of 1.0--2.5~eV for the two valleys, denoted as $I_K(t)$ and $I_Q(t)$, and track their time evolution. 
A phenomenological dephasing parameter is typically needed to account for the ultimate relaxation of the excitations for mechanisms not included in the adopted formalism~\cite{Chan2021,chang2024many,hu2023light}.
Here, as shown in Fig.~\ref{fig3}\textbf{b}, we include a dephasing time constant of $\tau_\text{deph} = 2.2$~ps, determined by fitting the decay of the experimentally measured intensity $I_K + I_Q$ to an exponential function (see SI)~\cite{Madeo2020}. 
Without dephasing, the calculated total excitation intensity remains essentially constant after the pump is turned off, dictated by the conservation of the total spectral weight.

Fig.~\ref{fig3}\textbf{c} and \textbf{d} show the direct comparison of the exciton dynamics in the \textit{K}-valley and \textit{Q}-valley, respectively, between the TD-a$GW$-ph calculation and experiment \cite{Madeo2020,Pareek2026}.
Remarkably, our calculated time-resolved intensities $I_K$ and $I_Q$ reproduce the experimental tr-ARPES data in the overall behavior and the timescale.
The agreement is further corroborated by the intensity ratio $I_K/I_Q$ as shown in Fig.~\ref{fig3}\textbf{e}, which decreases from $>$1.5 at early time to a quasi-equilibrium of $\sim$0.5 beyond $\sim$500~fs, suggesting that the dark $Q$-valley excitons become the dominant excitations at the longer time scale.
Note that the sub-100~fs oscillatory feature in Fig.~\ref{fig3}\textbf{c}-\textbf{e} directly reflects the coherent coupling between excitons and phonons and is a consequence of the single-trajectory simulation of the phonon dynamics. Under experimental conditions (e.g. large sample size with numerous domains), ensemble averages will smoothen out such fast oscillations, but the main characteristics of the exciton dynamics will survive as a result of phonon-induced broken translational symmetry (of the primitive unit cell).
We note that the ultrafast intervalley dynamics captured by TD-a$GW$-ph constitutes the main dephasing channel for the bright \textit{K}-valley excitons.
In the clamped-nuclei TD-a$GW$ simulation in Fig.~\ref{fig2}\textbf{b}, the same $\tau_\text{deph} = 2.2$~ps is applied and the \textit{K}-valley intensity of the direct exciton remains strong even at 1~ps, missing the ultrafast $\sim$200~fs dynamics as shown in Fig.~\ref{fig3}\textbf{a} and \textbf{c}.
In previous TD-a$GW$ simulations of excitons in transition-metal dichalcogenides, a much shorter dephasing time of 10--500 fs (with high variation)~\cite{Chan2021,chang2024many,hu2023excitonic} is required to qualitatively capture the ultrafast decay of excitons in the \textit{K}-valley. 
In this work, we show that most ($\gtrsim$80\% in terms of $1/\tau_\text{deph}$) of the dephasing processes of the \textit{K}-valley excitons can be attributed to the phonon-mediated \textit{K}-to-\textit{Q} intervalley exciton dynamics.

Besides the dynamics reflected in the intensity distribution across valleys, phonons directly affect the spectral line shape.  
Fig.~\ref{fig3}\textbf{f} and \textbf{g} show the energy line cuts of $I(E,t)$ in Fig.~\ref{fig3}\textbf{a} at representative times, which can be fitted by Lorentzian functions  (see SI).
Fig.~\ref{fig3}\textbf{h} summarizes the extracted Lorentzian half-width-at-half-maximum spectral width $d(t)$, and we find $d_K$ and $d_Q$ at both valleys essentially increase gradually after the pump pulse ends in the TD-a$GW$-ph simulations. The broadening of the spectral intensity suggests that the classical phonon perturbations induce variations in the instantaneous energy levels (in analogy to the configuration sampling using frozen-phonon-based approaches~\cite{zacharias2016one}), and the redistribution of the excitonic spectral weights over a broader energy and momentum range through both intervalley and intravalley scattering. 
Note that, however, the current classical treatment of phonons may overestimate phonon absorption at low temperature, exaggerating the scattering of excitons into higher-energy states and thereby contributing to the growth of the spectral width at longer times. A quantum treatment of the phonon statistics is expected to provide more accurate descriptions of the long-time spectral line shape.
On the other hand, our TD-a$GW$-ph results demonstrate that the ultrafast \textit{K}-to-\textit{Q} exciton dynamics can be directly mediated by thermal background phonons alone, highlighting the importance of incorporating phonon effects for describing realistic exciton dynamics that is inaccessible in the clamped-nuclei TD-a$GW$ approach.

\textit{Conclusion.}
In summary, we have developed an \textit{ab initio} TD-a$GW$-ph approach that incorporates coherent $e$-$h$ and $e$-ph couplings across the full BZ for real-time simulations of exciton-phonon coupled dynamics. This method well captures the ultrafast \textit{K}-to-\textit{Q} intervalley exciton transfer dynamics in monolayer WSe$_2$, showing excellent agreement with experiment. Our work paves the way to study complicated light-driven phenomena and processes, including exciton transfer pathways, self-trapped exciton formation, and the coherence of exciton and exciton-phonon coupled dynamics in solids.

\textit{Acknowledgment.}
Z.L. thanks Gianluca Stefanucci for insightful discussions.
This work was primarily supported by the Center for Computational Study of Excited-State Phenomena in Energy Materials at the Lawrence Berkeley National Laboratory, which is funded by the United States Department of Energy (DOE), Office of Science, Basic Energy Sciences, Materials Sciences and Engineering Division under Contract No. DE-AC02-05CH11231, as part of the Computational Materials Sciences Program, which provided the development of methods.
The work was partially supported by the U.S. National Science Foundation (NSF) under Grant No. OAC-2513830 for enabling interoperable software calculations (S.G.L. and Z.L.).
Z.L. and T.C. acknowledge the support by the U.S. NSF CAREER Awards under Grant No. DMR-2440763 and DMR-2339995, respectively, for the first-principles calculations and analyses.
Y.H.C. acknowledges the support of Academia Sinica under project No. AS-CDA-114-M04.
C.-E.H. and H.-C.H. acknowledge support from the National Science and Technology Council (NSTC), Taiwan, under Grant Nos. NSTC 113-2112-M-032-013, 114-2112-M-032-009-MY3, and 114-2811-M-032-012-MY2.
An award of computer time was provided by the U.S. DOE INCITE program.
This research used computational resources of both the Argonne and Oak Ridge Leadership Computing Facilities, which are U.S. DOE SC User Facilities supported under contracts DEAC02-06CH11357 and DE-AC05-00OR22725. 
Computational resources were also provided by National Energy Research Scientific Computing Center, which is a U.S. DOE SC User Facilities supported under contract DEAC02-05CH11231, and by Texas Advanced Computing Center, which is supported by U.S. NSF under Grant No. OAC-1818253.
Y.H.C. thanks the National Center for High-Performance Computing (NCHC) for providing computational and storage resources.

\bibliography{ref}

\end{document}